\documentclass[fleqn,usenatbib]{mnras}

\usepackage{newtxtext,newtxmath}
\usepackage[T1]{fontenc}

\DeclareRobustCommand{\VAN}[3]{#2}
\let\VANthebibliography\thebibliography
\def\thebibliography{\DeclareRobustCommand{\VAN}[3]{##3}\VANthebibliography}

\usepackage{graphicx}	% Including figure files
\usepackage{amsmath}	% Advanced maths commands
\usepackage{xcolor}
\def\wcen{{$\omega$ Cen}}

\def\Pmem{{$\text{P}_{\text{mem}}$}}
\def\kms{{\rm\,km\,s^{-1}}}

\def\masyr{{\rm\,mas\,yr^{-1}}}
\def\kpc{{\rm\,kpc}}

\def\msun{{\rm\,M_\odot}}

\def\s{\ifmmode \widetilde \else \~\fi}
\def\={\overline}

\def\spose#1{\hbox to 0pt{#1\hss}}

\def\lta{\mathrel{\spose{\lower 3pt\hbox{$\mathchar"218$}}
     \raise 2.0pt\hbox{$\mathchar"13C$}}}
\def\gta{\mathrel{\spose{\lower 3pt\hbox{$\mathchar"218$}}
     \raise 2.0pt\hbox{$\mathchar"13E$}}}
\def\Dt{\spose{\raise 1.5ex\hbox{\hskip3pt$\mathchar"201$}}}    % upper case
\def\dt{\spose{\raise 1.0ex\hbox{\hskip2pt$\mathchar"201$}}}    % lower case

\def\dotsfill{\leaders\hbox to 1em{\hss.\hss}\hfill}

\def\FeH{{\rm[Fe/H]}}

\title[Pristine view of \wcen]{A Pristine View of Galactic Globular Clusters and their Peripheries: Omega Centauri}

\author[P. B. Kuzma et al.]{
P. B. Kuzma,$^{1,2}$\thanks{JSPS International Research Fellow}\thanks{E-mail: pete.kuzma@nao.ac.jp}
M. N. Ishigaki,$^{1}$
\\
$^{1}$National Astronomical Observatory of Japan, 2-21-1 Osawa, Mitaka, Tokyo 181-8588, Japan\\
$^{2}$Institute for Astronomy, University of Edinburgh, Royal Observatory, Blackford Hill, Edinburgh, EH9 3HJ, UK\\
}

\date{Accepted XXX. Received YYY; in original form ZZZ}

\pubyear{2024}

\begin{document}
\label{firstpage}
\pagerange{\pageref{firstpage}--\pageref{lastpage}}
\maketitle

\begin{abstract}
The central regions of the globular cluster Omega Centauri (\wcen) have been extensively studied, but its outer regions and tidal structure have been less so. Gaia's astrometry uncovered substantial tidal substructure associated with \wcen, yet the lack of chemical tagging makes these associations tenuous. In this paper, we utilise the Gaia-Synthetic CaHK-band photometry, metallicities from the Pristine survey and Gaia's astrometry to explore up to a clustercentric radius of 5 degrees from \wcen. We identify \wcen-like stars based on proper motion, colour-magnitude and colour-colour space, exploring the morphology, and stellar populations of the outer regions. Our probabilistic approach recovers the tidal tails of \wcen, and we investigate the metallicity distribution of \wcen$\,$ranging from a radius of 15 arcmin to the tidal radius, and beyond into the tidal tails. We present (1) two components between 15 arcmin and tidal radius at -1.83 and 1.45 dex which are also the dominant populations within 15 arcmin, and (2) the first evidence that the same two populations in the outer regions of the cluster are present outside the tidal radius and into the tidal tails. These populations are mixed about the stream, and are typically amongst the faintest stars in our sample; all indicating that the tidal tails are made of tidally stripped \wcen$\,$stars.

\end{abstract}

%%%%%%%%%%%%%%%%%%%%%%%%%%%%%%%%%%%%%%%%%%%%%%%%%%

%%%%%%%%%%%%%%%%% BODY OF PAPER %%%%%%%%%%%%%%%%%%

\section{Introduction}
\label{sec:intro}
As the most massive globular cluster in the Milky Way, $\omega$ Centauri (NGC 5139) has received its fair share of attention. Its enigmatic nature largely hinges on its phenomenal stellar populations. Photometric studies have uncovered the presence of multiple stellar populations \citep[e.g.,][]{2000ApJ...534L..83P,2004ApJ...605L.125B,2005MNRAS.357..265S,2007ApJ...663..296V,2009A&A...507.1393B,2010AJ....140..631B,2017ApJ...844..164B,2017AJ....153..175C,2017MNRAS.469..800M}, and spectroscopic studies of giant stars within $\omega$ Centauri (hereafter \wcen) have shown many complicated features: from the light elemental abundance variations of Na-O, Al-O, and Mg-Al amongst others that are common in Milky Way (MW) GCs \citep[][and references therein]{1975ApJ...201L..71F,1995ApJ...447..680N,2010ApJ...722.1373J,2017ApJ...844..164B,2021MNRAS.505.1645M,2024A&A...681A..54A}, to the more unique heavy element abundance variations \citep[e.g.,][]{2003MNRAS.345..683P,2011ApJ...731...64M,2011A&A...534A..72G,2020AJ....159..254J,2021A&A...653L...8L,2023arXiv230902503N}. These features were just two of the many properties that is perplexing about \wcen. Other stand-out properties include its retrograde orbit \citep[][]{1999AJ....117.1792D,2021MNRAS.505.5978V} and distinct kinematic signatures between its multiple stellar populations \citep[e.g.,][]{1997ApJ...487L.187N,2018ApJ...853...86B,2020ApJ...898..147C}. Collectively, this has led to the suggestion that \wcen$\,$was not formed {\it in-situ}, but that it is the core of a long-defunct dwarf galaxy \citep[e.g.,][]{1999Natur.402...55L,2000A&A...362..895H,2000A&A...357..977C,2003MNRAS.346L..11B,2014ApJ...791..107V,2024arXiv241022479P}. As our technologies advanced, we could start to analyse beyond the central regions of \wcen, and its nature became more complicated.

%One elusive piece of evidence that could further the cause of an accreted scenario for \wcen$\,$is the existence of tidal structure.

To explore this connection, one could search for evidence in the GCs outer regions. The existence of cluster-like stars outside a nominal radial boundary where the gravitational influence from the MW is stronger than that of the cluster, such as the observationally motivated tidal radius \citep[e.g.,][]{1962AJ.....67..471K} or the dynamically motivated Jacobi radius \citep[e.g.,][]{2001ASPC..228...29H}, may indicate the presence of some tidal stream or structure. However, in the case of \wcen, tidal structure was proving difficult to detect. Wide field photometry of \citet[][]{2000A&A...359..907L} would find compelling evidence for extended tidal structure in the form of tidal tails, but \citet[][]{2008AJ....136..506D} noted that there were no signs of extra-tidal stellar populations in the periphery through wide field spectroscopy. Definitive evidence for extended stricture surrounding \wcen$\,$would come when the Gaia Space Mission (Gaia) published its second data release \citep[][]{2016A&A...595A...1G,2018A&A...616A...1G}. Gaia, now on their third data release \citep[][]{2023A&A...674A...1G}, provided precision astrometry including on-sky motions and parallax measurements which allowed for superior methods of discarding contaminating MW field stars from GC stars to be developed. This is particularly important when searching for typically low surface brightness features hiding on the fringes of the overwhelming illumination of central regions. Several works utilising Gaia uncovered extra-tidal structure around many MW GCs \citep[][]{2019MNRAS.489.4565K,2019ApJ...884..174G,2019MNRAS.484L.114K,2019MNRAS.486.1667C,2020MNRAS.499.2157C,2020MNRAS.495.2222S,2022ApJ...929...89G,2023ApJ...953..130Y}, and \wcen$\,$ is no exception. \citet[][]{2019NatAs.tmp..258I} uncovered a tidal stream which they tagged to \wcen$\,$named {\it Fimbuthal} using their state-of-the-art STREAMFINDER algorithm \citep[][]{2018MNRAS.477.4063M} which identifies groups of stars based on their orbits, regardless of their spatial distribution on the sky. \citet[][]{2021MNRAS.507.1127K} also uncovered the tidal tails of \wcen, but through an unrelated method which utilized a mixture-model approach within a Bayesian framework. \citet[][]{2020MNRAS.495.2222S} also independently detected the tidal tails, enforcing the power of Gaia. Most recently, \citet[][]{2024ApJ...967...89I} claimed to have located a further extension of the trailing arm of the tidal stream of \wcen, denoted as {\it Stream $\#$55}. While both streams connect to \wcen$\,$through orbital kinematics, they do lack overall chemistry and line-of-sight kinematics to develop firm connections from the main body of \wcen, particularly as they are not spatially connected to \wcen, but are located along the same orbital path.

Gaia also provided insight into the proposed accreted nature of \wcen. Before Gaia, the largest and only known accretion event was the Saggitarius dwarf galaxy \citep[][]{1995MNRAS.277..781I}, and multiple GCs have been suggested to be brought into the MW halo with it \cite[e.g.,][]{1995AJ....109.2533D,2003AJ....125..188B,2010ApJ...718.1128L,2015A&A...579A.104S,2020A&A...636A.107B,2022ApJ...926..107M}. Since the release of Gaia DR2, many accretion events have been suggested to have taken place over MWs lifetime \citep[e.g.,][]{2019A&A...625A...5K,2020MNRAS.498.2472K,2022ApJ...926..107M,2022ApJ...930L...9M,2024ApJ...964..104M}. However, the most notable dwarf galaxy of them all is the Gaia-Sausage-Enceladus merger \cite[hereafter GSE;][]{2018MNRAS.478..611B,2018Natur.563...85H}. \citet[][]{2019A&A...630L...4M} identified \wcen$\,$as the core of GSE through connections in action energy space \citep[see also][]{2022ApJ...935..109L}. Intriguingly, this is not the only accretion event that \wcen$\,$has been linked to. The dwarf galaxy Sequoia, identified as a stand-alone accretion event by \citet[][]{2019MNRAS.488.1235M}, is another dwarf galaxy that places \wcen$\,$as its core. The picture is unclear as to which dwarf galaxy \wcen$\,$belongs to, or even another as-yet-detected accretion event. However, the proposed connection to either the accreted galaxies GSE or Sequoia remains difficult to ascertain given the troubles of linking GCs to accretion events through action-energy space \citep[][]{2023A&A...673A..86P}.

Chemically speaking, the notion of \wcen$\,$being the progenitor of the {\it Fimbuthal} stream and {\it Stream $\#$55}, or even the role of \wcen$\,$in either GSE or Sequoia remains unclear. A small number of spectroscopic works attempt to tag stars as tidally stripped stars through \FeH$\,$metallicities do provide tentative evidence that debris relating to \wcen$\,$can be found throughout the galaxy \citep[][]{2020MNRAS.491.3374S,2023MNRAS.524.2630Y}. In fact, shards linked to \wcen$\,$were found throughout the halo \citep[][]{2018MNRAS.478.5449M}. However, these connections remain tentative without the chemical connection to what stars are being stripped immediately from \wcen. In this work, we look to search for the \FeH$\,$content of stars in the periphery of GCs for the first time by utilising the Pristine survey's first data release \citep[][]{2023arXiv230801344M}. In the following sections, we will identify \wcen-like stars in the periphery through a probabilistic approach, and compare the stars in the periphery to the main body, and compare those to the external streams, and the GSE and Sequoia events, before presenting our concluding comments.

\begin{table}
\caption{Table of parameters for \wcen. First column denotes the parameters, second the corresponding value, and the third belonging source as follows: (1) \citet[][]{2021MNRAS.505.5978V}, (2) \citet[][]{2021MNRAS.505.5957B}, (3) \citet[][]{2018MNRAS.478.1520B}, (4) \citet[][]{2024ApJ...977...14C}, (5) \citet[][]{2019MNRAS.485.4906D}, (6) \citet[][]{2018MNRAS.474.2479B}.}
\centering
\begin{tabular}{||l l l||} 
\hline
\multicolumn{3}{|c|}{\wcen$\,$Parameters}\\
\hline\hline
$\alpha$ (J2000)& 13 26 47.28&(1)\\ 
$\delta$ (J2000) & -47 28 46.1&(1) \\
Dist.$_{\odot}$($\kpc$) & 5.43&(2) \\
Mass ($\msun$) & $3.94 \times 10^{6}$&(3)\\
\FeH$\,$(dex) & -1.7&(4) \\
$\text{V}_{rad}$ ($\kms$)& 232.78&(1)  \\ 
$\mu_{\alpha}$ ($\masyr$)& -3.236&(1) \\ 
$\mu^{*}_{\delta}$ ($\masyr$) & -6.731&(1)  \\
$\text{r}_{\text{core}/\text{c}}$ (arcmin)& 2.7&(5)\\
$\text{r}_{\text{tidal}/\text{t}}$ (arcmin)& 48.4&(5) \\
$\text{r}_{\text{Jacobi}/\text{j}}$ (arcmin)& 106.9&(6) \\ 
\hline
\end{tabular}
\label{tab:wcen_stats}\\

\end{table}

\begin{figure*}
    \centering
    \includegraphics[width=5in, height=5in]{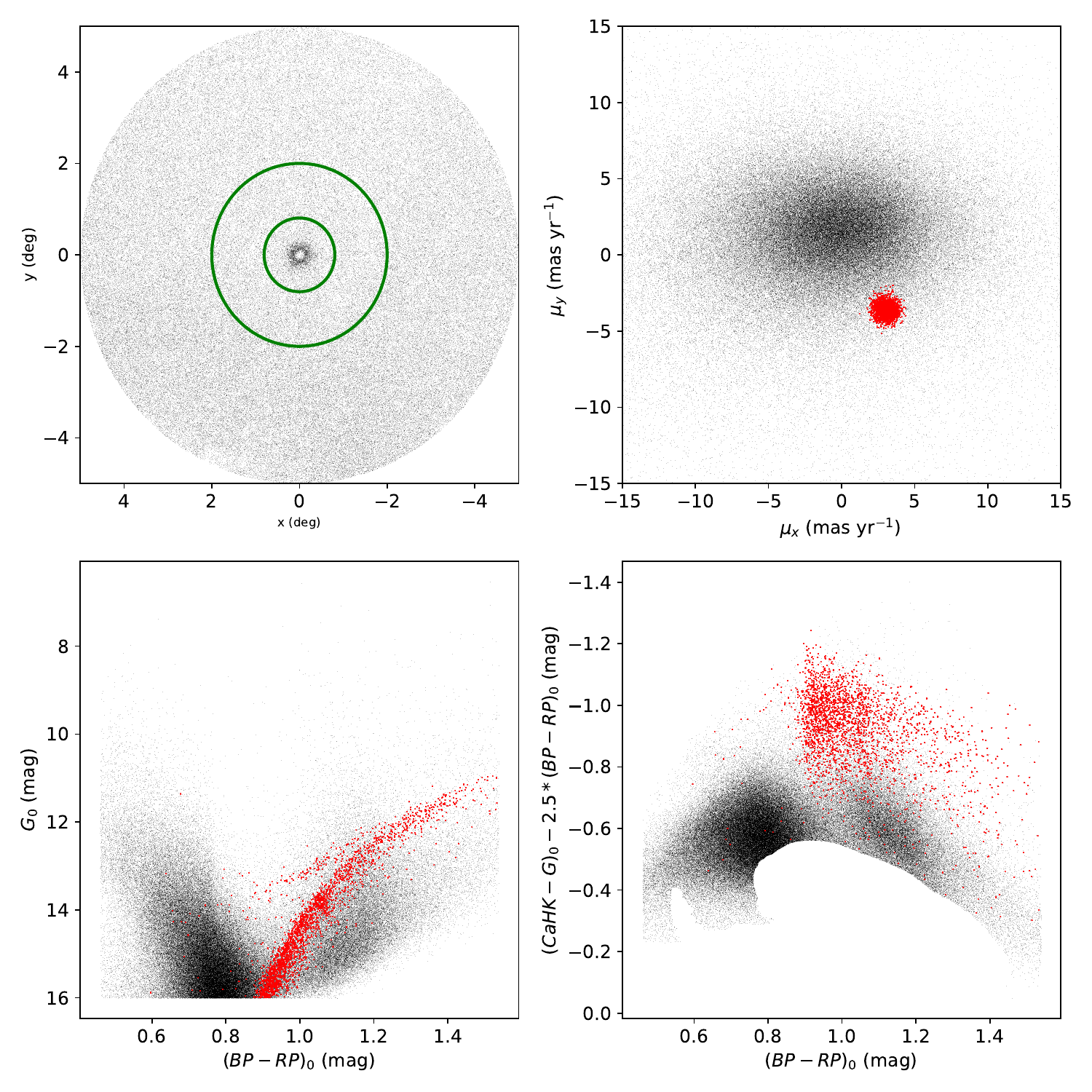}
    \caption{Top left: spatial distribution of all the retrieved stars in the Pristine-Gaia-synthetic catalogue. The two green circles indicate the cut-off radii for the {\it cluster} ($r_t$, 48 arcmin) and {\it field} (2 degrees) reference samples. Top right: Proper motion distribution of the {\it cluster} sample (red) and {\it field} reference sample (black.). The bottom row figures show the reference samples in colour-magnitude space (bottom left) and colour-colour space (bottom right) in the same colour scheme as in the top right figure.}
    \label{fig:CMD_CC_PM}
\end{figure*}

\section{Data Acquisition}\label{sec:data_acq}
This work utilises the Pristine-Gaia-synthetic catalogue of the first data release of Pristine\footnote{\url{https://sites.google.com/view/pristinesurvey/home?authuser=0}} \citep[][]{2023arXiv230801344M}. The synthetic catalogue provides photometric \FeH$\,$measurements and dereddened photometry in the form of Gaia passbands of $\rm{G_0}$, $\rm{G_{BP,0}}$ and $\rm{G_{RP,0}}$ along with the Pristine-unique $\rm{CaHK}$-band. Additionally, each member is cross-matched with their corresponding {\it source\_id} in Gaia DR3. Combining the metallicities, dereddened photometry and proper motions from the two surveys provides the perfect opportunity to search for tidally stripped in the peripheries of MW GCs. In the work, we will focus on \wcen$\,$(Table \ref{tab:wcen_stats}). We retrieved stars lying within a clustercentric radius of 5 degrees about the center of \wcen$\,$and cross-matched those stars with their Gaia DR3 counterparts. The resulting cross-matched catalogue is then subjected to a series of quality cuts that provide both data sets' highest-quality results. In particular, we retained stars in our cross-matched catalogue that met the following conditions \citep[see sec 7.3 of][]{2023arXiv230801344M}:
\begin{itemize}
\item{Measured metallicity in the range of $-4 <$ \FeH$\,$$< 0$, as measurements outside this range are automatically assigned $0$ or $-4$;}
\item{Uncertainty on \rm{\rm{[Fe/H]}} such that $\sigma_{\FeH}<0.3$ dex;}
\item{$P_{\rm{\rm{var}}} < 0.3$, removing any potential variable stars;}
\item{Typical Gaia cuts based on the Renormalised Unit weight error (\texttt{RUWE}  $< 1.4$) and corrected flux excess, $|C^{*}| < 1 \times 3\sigma_{C^{*}}$ which removes poor photometric and astrometric measurements from Gaia \citep{2018A&A...616A...2L,2021A&A...649A...2L};}
%\item{\texttt{\texttt{CASU\_flag}} $= -1$ or $-2$, which ensure the targets from Pristine are not extended objects or any other issues uncovered;}
\item{Stars whose parallax places them at a heliocentric distance beyond 3 kpc within uncertainties.}
\end{itemize}

Once all these quality cuts have been applied, we performed a series of coordinate transformations which transforms the spatial coordinate system ($\alpha$,$\delta$) and their associated proper motion ($\mu^{*}_{\alpha}$,$\mu^{*}_{\delta}$)\footnote{$\mu^{*}_{\alpha}=\mu_{\alpha} \cos{\delta}$} to a gnomonic tangential coordinate system $(x,y)$ and $(\mu_{x},\mu_{y})$, removing projection effects that may occur at large clustercentric radii, using the equations provided in \citet[][]{2018A&A...616A..12G}:
\begin{equation}
\begin{aligned}
x & = \cos\delta\sin(\alpha-\alpha_\text{0})\\
y & = \sin\delta\cos\delta_\text{0}-\cos\delta\sin\delta_\text{0}\cos(\alpha-\alpha_\text{0}) \\
\mu_{x} & = \mu_{\alpha}^{*}\cos(\alpha-\alpha_{\text{0}})-\mu_{\delta}\sin\delta\sin(\alpha-\alpha_{\text{0}})\\
\mu_{y} &= \mu_{\alpha}^{*}\sin\delta_\text{0}\sin(\alpha-\alpha_\text{0})\\
& \quad+\mu_{\delta}\left(\cos\delta\cos\delta_\text{0}+\sin\delta\sin\delta_\text{0}\cos(\alpha-\alpha_\text{0})\right)
\end{aligned}%\quad\right\}
\label{eq:tangent}
\end{equation}
\noindent

The uncertainties and covariances are all propagated appropriately. Additionally, we correct for solar reflex motion in the proper motions by adopting the heliocentric distance to \wcen$\,$and the solar motion $(12.9, 245.6, 7.78)$ $\kms$ \citep{2018RNAAS...2..210D}. Lastly, catalogue homogeneity is sensitive not only to the scanning law of Gaia that Pristine cross-matched to but to the level of extinction across the field of view. The region being explored here is subject to small variable extinction across the field view, particularly in the direction of the Galactic Center. These variations across the field of view can have a negative effect on identifying any co-moving stars in over-densities that may be a result of varying extinction, such as different regions of the field of view having different photometric depths. Further, the residual scanning pattern from Gaia also affects the appearance of overdensities. These patterns are not uniform across the field of view \citep[e.g., see Fig 10 of][]{2023A&A...669A..55C}, and the photometric depth can change depending on how many passes Gaia has made in the field of view. All in all, these can create overdensities that are artificial artefacts, rather than true groupings of GC stars. To minimize the effects of reddening we apply a photometric cut at $\text{G}_0$$=$ 16 mag, removing all stars fainter than this limit.

\section{Methods}\label{sec:method}
Armed with the combined photometry and astrometry from Pristine and Gaia, we aimed to develop a probabilistic technique which can utilise multi-band photometry, particularly the CaHK-band filter, to help identify tidal stripped stars in the periphery of \wcen. The CaHK-band filter is sensitive to metallicity and therefore is a useful discriminator from the contaminating field. This is particularly important for \wcen, due to its Galactic latitude. Combine this filter with the photometry of astrometry of Gaia, and we have a multi-dimensional space to explore. Using \citet[][]{2021MNRAS.507.1127K} as inspiration, we assess the probability of a star belonging to either the field or \wcen$\,$in the 10-degree diameter field of view by exploring the proper motion distribution, location in the colour-magnitude (CMD) space, and location in colour-colour (CC) space. We started by creating two reference sets which we denote {\it cluster} and {\it field} (shorthand {\it cl} and {\it MW} respectively). The {\it cluster} set is defined as a subset of our combined catalogue that best describes the main body of the \wcen, and the reference set contains all stars within the tidal radius of \wcen$\,$(48 arcmin), and all stars within proper motions within 1.5 $\masyr$ of the bulk proper motion of \wcen. As for the {\it field} reference sample, we selected all stars outside a clustercentric radius of 2 degrees. In Fig. \ref{fig:CMD_CC_PM}, we demonstrate all stars in Pristine-Gaia \footnote{We note that the lack of completeness in the central regions in Fig. \ref{fig:CMD_CC_PM} are a product of completeness crowding within Gaia DR3. A crossmatch of Pristine-Gaia catalogue and the Focus Data Product Release of Gaia \citep[][]{2023A&A...680A..35G} was not performed as part of Pristine DR1.}, and we compare the two samples in each of the three spaces, and it is clear that the cluster stands out against the field: the cluster is separated from the field in proper motion space, and the red giant branch (RGB) stands-out clearly in the CMD. The multiple stellar populations of \wcen$\,$are also readily apparent, particularly in the colour-colour diagram due to the large vertical extent of the CMD. These two samples will help define the analytical models that will calculate the membership probabilities $P^{cl}$ and $P^{MW}$ for the cluster members and the MW field contaminates respectively.
The likelihood of the proper motion for both reference sets is modelled as a 2-dimensional normal distribution:
\begin{equation}
\ln{P_\textrm{PM}}=-\frac{1}{2}(X-\bar{X})^\top\,C^{-1}\,(X-\bar{X}) - \frac{1}{2}\ln{4\pi\det[C]} 
\label{eq:lpm}
\end{equation}
\noindent
where $C$ is the covariance matrix:
\begin{equation}
C=\begin{bmatrix}
\sigma^2_{\mu_{x}}+\sigma^2_{x}& p\sigma_{\mu_{x}} \sigma_{\mu_{y}}\\
p\sigma_{\mu_{x}} \sigma_{\mu_{y}}& \sigma^2_{\mu_{y}}+\sigma^2_{y}\\
\end{bmatrix} 
\label{eq:cov}
\end{equation}
\noindent
where $X$ = ($\mu_{x},\mu_{y}$) are the proper motions of the reference set stars, ($\sigma^2_{\mu_{x}},\sigma^2_{\mu_{y}}$) is the corresponding uncertainties on the proper motions, $p$ is the covariance, $\bar{X}$ is the bulk motion and ($\sigma_{x}$,$\sigma_{y}$) is the proper motion dispersion of the cluster/field.

The next space to explore is the dereddened Gaia colour-magnitude space of ($\rm{G_{BP,0}-G_{RP,0},G_0}$). Here we adopt local density for any given star in the full sample, with respect to each reference sample. A star that lies close to the RGB of \wcen$\,$ will more likely be a cluster member than those lying further away, in and amongst the field population. Therefore, we utilise $k$-nearest neighbours approach. We looked to find the $k=10$ nearest neighbours for a given star in both reference samples and calculate the local density. In our technique, we aim to utilise the uncertainties in colour and magnitude, but the typical relative uncertainty in magnitude is smaller than in colour. Therefore, to appropriately consider these differences in sizes of uncertainty and the effects that may have on our algorithm, we calculated the average ratio in the photometric uncertainties between colour and magnitude in the cluster reference sample \citep[see also][]{2020MNRAS.495.2222S}. That ratio was found to be 1:2. We reflect this ratio as a metric in our probability equation for the CMD local density, and that equation takes the following form:

\begin{equation}
\ln P_{\rm{\rm{CMD}}} = \ln(\frac{10}{\pi (\Delta^2_{\rm{G_{BP,0}-G_{RP,0}}}+(\frac{\Delta_{\rm{G_0}}}{2})^2)})-\ln(N)  \label{eq:pcmd}
\end{equation}

Where $\Delta_{\rm{G_{BP,0}-G_{RP,0}}}$ and $\Delta_{\rm{G_0}}$ is the difference in colour and magnitude respectively between the star in question and its 10th nearest neighbour, and $N$ is the total number of stars in each reference sample.

While the two former spaces are commonly used to search for kindred stars, the colour-colour space is unique to Pristine. This space helps discern populations of different metallicities based on their photometric properties, which is particularly useful with metal-poor populations such as those seen in GCs. The membership probability in colour-colour space ($P_{\rm{CC}}$), the space of colour index $\rm{(CaHK - G_{0}) - 2.5 (G_{BP,0}-G_{RP,0})}$ (hereafter $\rm{CaHK_{ind}}$) with respect to the Gaia colours, take the same form as eq. \ref{eq:pcmd}. However, here the metric between the two colour indexes is formulated by finding the mean ratio between the typical uncertainties between the $\rm{(G_{BP,0}-G_{RP,0})}$ and $\rm{CaHK_{ind}}$ colours. In this case, the ratio is 10:1. Therefore, eq. \ref{eq:pcmd} becomes:

\begin{equation}
\ln P_{\rm{CC}} = \ln(\frac{10}{\pi ((\frac{\Delta_{\rm{(G_{BP,0}-G_{RP,0}})}}{10})^2+\Delta_{\rm{CaHK_{ind}}}^2)})-\ln(N)\label{eg:pcc}
\end{equation}
where everything is the same as in eq. \ref{eq:pcmd} except that $\Delta_{\rm{CaHK_{ind}}}$ is the difference in the colour index of the star in question and its 10th nearest neighbour.
With each membership probability now defined, we aimed to find the relative ratio of cluster-to-field stars, denoted as $f_{cl}$, for the field of view explored. This is done by maximising the following likelihood function:
\begin{equation}
L = f_{cl} P^{cl}_{PM}P^{cl}_{CMD}P^{cl}_{CC} + (1-f_{cl})P^{MW}_{PM}P^{MW}_{CMD}P^{MW}_{CC}
\end{equation}
where $cl$ refers to the cluster and f refers to the MW field in the field of view. Once $f_{cl}$ is identified, we assign each star in the field a value, $P_{\mathrm{mem}}$  between 0 and 1, where 1 is the highest probability that a target is a member of \wcen$\,$in the field of view, and 0 suggests the target is a member of the field based on their proper motions, and location in both CMD and CC space. The assignment follows the form:
\begin{equation}
P_{\mathrm{mem}} = f_{cl} P^{cl}_{PM}P^{cl}_{CMD}P^{cl}_{CC}/L
\end{equation}
It is with $P_{\mathrm{mem}}$ that we will use moving forward as we explore the peripheries of \wcen. 
Before exploring the resulting probabilities in each parameter space previously described, we created a 2D density map demonstrating the regions in the field with an over-density of high-probability stars. Following the approach of \citet[][]{2020MNRAS.495.2222S}, we placed a grid of width 6 x 6 arcmin across the field of view between the boundaries of +/-5 degrees in both the $x-$ and $y-$ directions. At each intersection, we calculated the local density of stars:
\begin{equation}
\rho(x,y) = \frac{N_{xy}}{\pi R^2_{loc}}
\end{equation}
where $N_{xy}$ is the number of all neighbouring stars that satisfy the condition that the sum of $P_{\mathrm{mem}}$ are at least 2, and $R_{loc}$ is the largest radius when the previous condition is met. Adopting this technique allows for an adaptive smoothing across the field of view. That is, regions with a high number of high probability stars will present a higher density value with a small smoothing kernel, such as within the tidal radius \wcen, while allowing for a larger smoothing over regions where there is an overall lower density of potential members. Our technique, while providing every star with a membership probability, still has a chance of underlying field distribution affecting any over-densities we may detect. Therefore, we fit a 2D first-order polynomial to the density map, excluding a clustercentric radius of 2 degrees (the distances where we deemed the overall population of \wcen$\,$to be significantly out-populated by the field) and subtract it from the original density map. Lastly, we calculate the average bin density $\overline{\rho}$ across the field of view and its associated standard deviation $\sigma_{\rho}$, excluding the inner two degrees again. Excluding the inner two degrees prevents the overwhelming density of the central regions of \wcen$\,$from affecting the fainter peripheries. Once the mean bin density and the associated standard deviation are found, we present each bin as $(\rho - \overline{\rho})/\sigma_{\rho}$, the number of standard deviations above or below the mean bin density.

\begin{figure*}
    \centering
    \includegraphics[width=\textwidth]{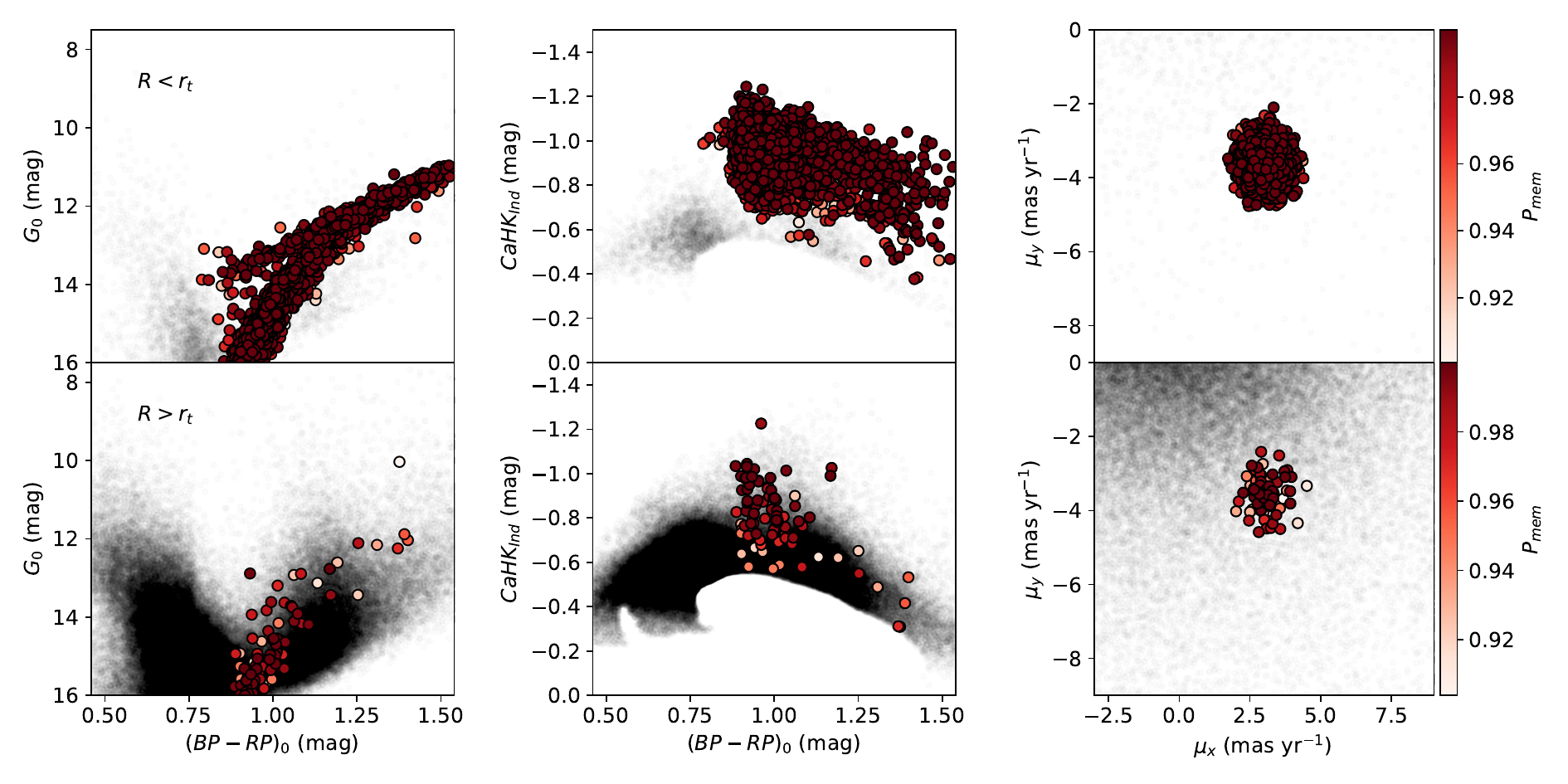}
    \caption{Distributions across the three-parameter spaces shown in Fig. \ref{fig:CMD_CC_PM} of stars with higher membership probabilities $P_{\mathrm{mem}} > 0.9$, coloured by their $P_{\mathrm{mem}}$ value, and stars with $P_{\mathrm{mem}} < 0.1$ in black. From left to right, the parameter spaces are CMD space, colour-colour space (middle) and proper motion space. The top row shows stars within the tidal radius (48.4 arcmin) and the bottom row demonstrates stars outside the tidal radius. In each figure, the grouping of high-probability stars closely follows the same features in Fig. \ref{fig:CMD_CC_PM}.}
    \label{fig:SPACE_prob}
\end{figure*}

\begin{figure}
    \centering
    \includegraphics[width=\columnwidth]{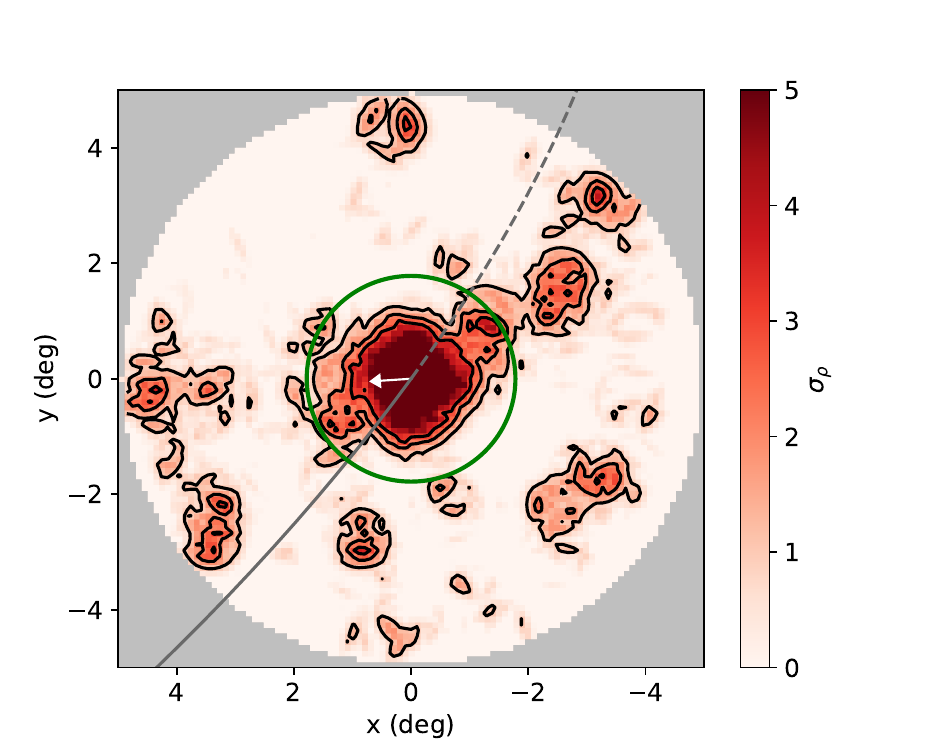}
    \caption{Surface density 2D distribution of all stars in our sample using the variable smoothing technique. Each star has been weighted by its membership probability. Colours relate to the number of $\sigma$ (1,2 and 3) above the mean bin value outside a clustercentric radius of 2 deg. The forward (backward) orbit \citep[][]{2015ApJS..216...29B} is shown by the solid (dashed) grey line, the Jacobi radius (106.9 arcmin) is indicated by the green ring, and the white arrow is pointing towards the Galactic Centre. The tidal structure is seen along the orbit.}
    \label{fig:2d_prob}
\end{figure}

\section{Results}\label{sec:results}
\subsection{2D Density Distribution}
We display the probabilities of two samples, calculated in the spaces of proper motion, CMD and CC-index in Fig. \ref{fig:SPACE_prob}. The samples, one which demonstrates stars with higher probabilities $P_{\mathrm{mem}}>0.9$ coloured by their membership probability, and the other with $P_{\mathrm{mem}}<0.1$, are split into two radial regions, one within $r_t$ (top row), the other outside $r_t$ (bottom row). Across the three-parameter spaces, the higher probabilities are all seen to closely follow the cluster sample as seen in Fig \ref{fig:CMD_CC_PM}, as do the stars outside the tidal radius. The broad distribution in $\rm{CaHK_{ind}}$ follows the broad metallicity distribution of \wcen, and the broad RGB of \wcen$\,$is seen in the CMD distribution. Lastly, the proper motion distribution of \wcen$\,$has a peak at ($3.05 \pm 0.01$,$-3.59 \pm 0.01$) mas yr$^{-1}$, calculated using the cluster sample and eq. \ref{eq:lpm}, is equivalent to the bulk motion of the cluster in \citet[][]{2021MNRAS.507.1127K}, corrected for solar reflex motion. Transforming the proper motion back with the inclusion of solar reflex motion, the bulk cluster motion is (-3.234, -6.719) mas yr$^{-1}$, which agrees with \citet[][]{2021MNRAS.505.5978V}. Collectively, our technique identifies similar in metallicity and kinematically coherent stars across the field of view, and this is also demonstrated by the 2D surface density map presented in Fig. \ref{fig:2d_prob}. Fig. \ref{fig:2d_prob} shows the density regions of high statistical significance, indicated by the colours and accompanying contour levels of 1, 2 and 3$\sigma$. Accompanying the 2D map is a demonstration of the orbit of \wcen, which is based on the 3D kinematics and distance presented in Table \ref{tab:wcen_stats}, and integrated with the \texttt{MWPotential2014} Galactic potential from \citet[][]{2014ApJ...795...95B}. The 2D map of stars weighted by their membership probabilities (see \S \ref{sec:method}) demonstrates clear elongation to the Jacobi radius, with several over-densities near the orbit that are consistent with the orientation of the tidal tails. The overall shape and direction of the debris is very similar to that uncovered in \citet[][]{2021MNRAS.507.1127K}, though a few notable differences stand out. The trailing tail remains detectable, but we do lose some of the leading tail (in the southeast direction) in our detection. Though this is most likely due to the photometric depth covered here when the previous studies are three to four magnitudes deeper and reach the main sequence of \wcen. Furthermore, our technique relies more on colour-colour spaces exploration and makes no attempt to model the underlying spatial distribution like \citet[][]{2021MNRAS.507.1127K}, chiefly due to the much shallower photometric depth that the Pristine-Gaia-synthetic catalog reaches. Regarding the axis of the debris, we measure the position angle $\theta$ of the extensions at the Jacobi radius by isophotal fitting to be $\theta = 117 \pm 3$ deg, which is in firm agreement with 122 deg presented in \citet[][]{2021MNRAS.507.1127K}. The angle of elongation and many of the over-densities are consistent with tails presented by \citet[][]{2019NatAs.tmp..258I}, \citet[][]{2020MNRAS.495.2222S} and \citet[][]{2021MNRAS.507.1127K}, though we do see elongation that is slightly misaligned with the orbit. The orbit computed implies that \wcen$\,$ has recently passed apogalacticon and is commencing an approach towards the disk of the MW. \citet[][]{2015MNRAS.446.3100H} demonstrated that as a satellite cluster approaches apogalacticon about its orbit, newly stripped stars will be preferentially pointing towards the Galactic centre, appearing as perpendicular tidal tails to the orbit. As the satellite passes apogalacticon, those newly stripped stars will begin to follow the orbit again.  According to the orbit of \wcen$\,$derived and shown in Fig \ref{fig:2d_prob}, \wcen$\,$ recently passed apogalacticon and is commencing its approach towards the orbit pericenter. The position angle and the shape of the debris and many fragments are consistent with this view of the development of tidal tails. Ultimately, we have demonstrated that the tidal tails of \wcen$\,$ are detectable with Pristine. 

\subsection{Metallicity Distributions}
\begin{figure*}
    \centering
    \includegraphics[width=\textwidth]{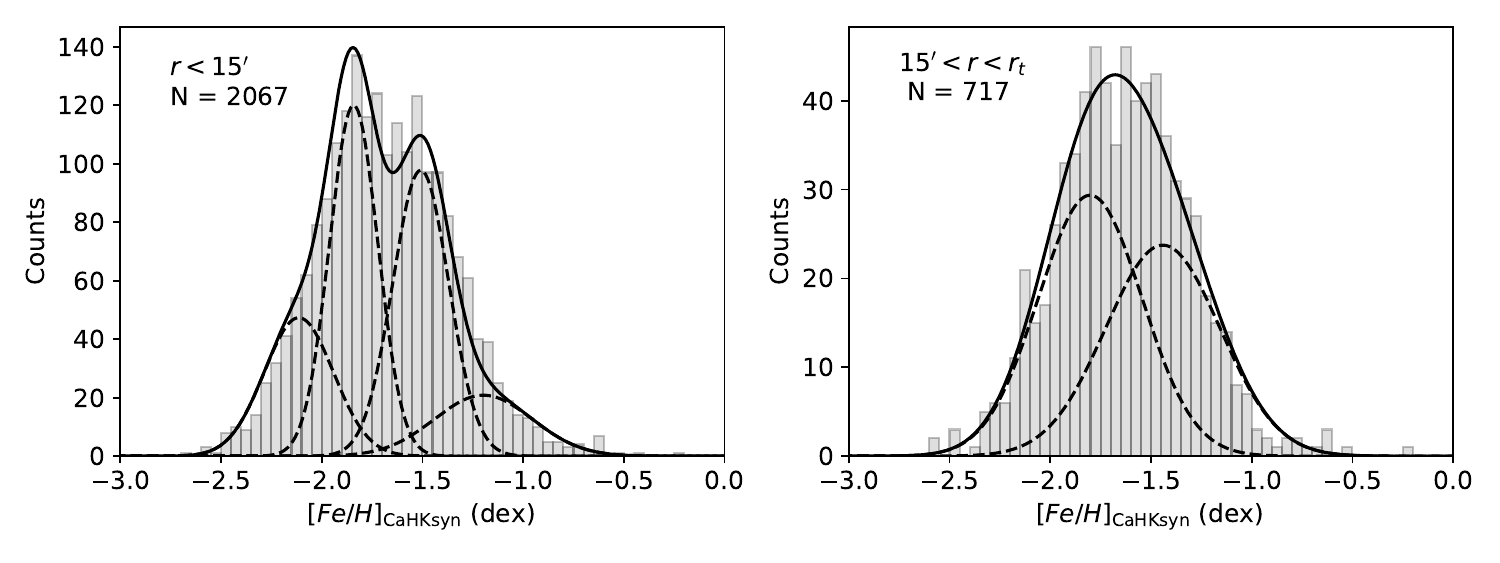}
    \caption{Metallicity distributions of all stars within 15 arcmin (left), and between 15 arcmin and $r_t$ (right). The solid black line shows the best fit 1D-Gaussian mixture model, with each dashed line identifying the underlying distributions (peaks at $-2.11 \pm 0.02$, $-1.83\pm0.02$, $-1.50\pm0.02$ and $-1.22\pm0.05$ dex on the right, and $-1.82\pm0.03$ and $-1.45\pm0.03$ dex for the left).}
    \label{fig:1d_hist_CL}
\end{figure*}

Despite the loss of stars in the central regions of \wcen\;due to crowding effects, many stars are still available to quantify metallicity distribution. The spread in $\rm{CaHK_{ind}}$ demonstrated in Fig. \ref{fig:CMD_CC_PM} demonstrates that Pristine should be able to identify some of the stellar populations of \wcen. We proceeded by sorting stars within the tidal radius into two radial groups - one that covers the inner 15 arcmin, and another that contains stars within a radial range of 15 arcmin to $r_t$. We fit a 1D Gaussian mixture model to the metallicity distributions of these two groups and used Akaike Information Criterion fitting  \citep[AIC;][]{1311138} to assess the appropriate number of 1D Gaussian distributions that best describe the data. The metallicity distributions and corresponding best fit 1D Gaussian Mixture model is shown in Fig. \ref{fig:1d_hist_CL}, which contains a histogram of both groups of bin widths of 0.05 dex. According to our AIC findings, the metallicity distribution of the inner 15 arcmin is best fit by four Gaussian distributions, which are indicated in the left plot of Fig. \ref{fig:1d_hist_CL} by the dashed lines. The identified stellar populations have peaks at $-2.11 \pm 0.03$, $-1.83 \pm 0.02$, $-1.50\pm0.02$ and $-1.19\pm0.06$ dex. As for the 2nd group between 15 arcmin and the tidal radius, we identify two populations at $-1.82 \pm 0.03$ and $-1.45 \pm 0.03$ dex. The uncertainties are estimated by repeating the above 1000 times using a bootstrapping method with replacement. Fig. \ref{fig:1d_hist_CL} demonstrates that the most dominant population is near -1.8 dex, and the second most at -1.5 dex. These two populations are also present in the outer radial range explored as the only detectable populations. 
The peaks of our distribution are mostly consistent with findings from other studies:
\begin{itemize}
    \item \citet[][]{2005ApJ...634..332S} presented four populations at \FeH$\,=-1.7, -1.3,-1.0$, and $-0.6$ dex based on with VLT/FLAMES spectroscopy;
    \item \citet[][]{2010ApJ...722.1373J} found five populations at \FeH$\,=-1.75, -1.50, -1.15, -1.05$, and $-0.75$ dex based on Hydra multifiber spectroscopy;
    \item \citet[][]{2014ApJ...791..107V} identified six populations at \FeH$\,=-1.83, -1.65, -1.34, -1.05, -0.78$ and $-0.42$ dex based on VLT/FLAMES spectroscopy;
    \item \citet[][hereafter M21]{2021MNRAS.505.1645M} published four populations at \FeH$\,=-1.65, -1.35,-1.05$ and $-0.7$ dex (including a +0.1 dex offset, see section 3 in M21) with the BACCHUS code on APOGEE DR16 spectra.
\end{itemize}
The most prominent population across these works are around \FeH$\,$of -1.8 or -1.7 dex, which is consistent with our findings here in both radial ranges explored. We are unable to detect significant populations larger than -1.0 dex that are present in the studies mentioned above, but Fig. \ref{fig:1d_hist_CL} does show a small number of stars above -1.0 dex, just not enough in number to make a firm detection. None of the above studies detect a population that is similar to the most metal-poor population in the Pristine-Gaia-synthetic catalogue near -2.1 dex. \citet[][]{2020AJ....159..254J} does identify a metal-poor population in \wcen, though it is mostly centrally concentrated (within 5 arcmin). However, the existence of this population may be due to crowding affecting the Gaia XP spectra, which would in turn affect the synthetic photometry of the Pristine-Gaia-synthetic catalogue (see \S \ref{sec:Disc_1}).

\begin{figure}
    \centering
    \includegraphics[width=\columnwidth]{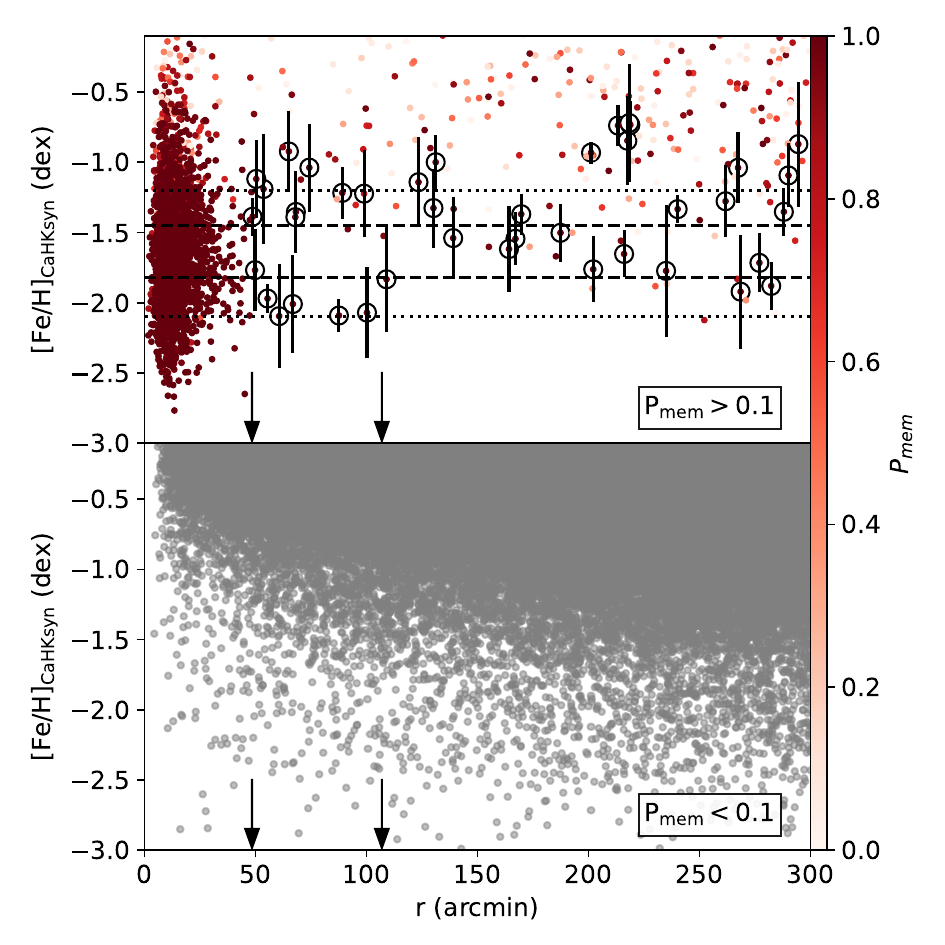}
    \caption{\FeH$_\textrm{CaHKsyn}\;$against radius in arcmin. Top: All stars with \Pmem > 0.1. Each star is coloured by its membership probability. Horizontal lines indicate the peaks of the metallicity distribution presented in Fig. \ref{fig:1d_hist_CL}: the dashed lines are the two populations detected between 30 arcmin and the tidal radius, and the dotted lines are the other two populations identified within the tidal radius. The vertical arrows are the tidal and Jacobi radii. The conservative sample stars $(P_{\mathrm{mem}}>0.99)$ are denoted by the white rings. The multiple populations are prominent, particularly between the tidal and Jacobi radii. The error bars on the conservative sample represent the uncertainties in \FeH$_\textrm{CaHKsyn}$ from the Pristine-Gaia-synthetic catalogue. Bottom: The distribution of stars \Pmem<0.1. The field distribution indicates the potential effects of residual contamination seen in the top plot at increasing clustercentric radii, even at modest levels of probabilities.}
    \label{fig:R_V_FeH}
\end{figure}

\begin{figure}
    \centering
    \includegraphics[width=\columnwidth]{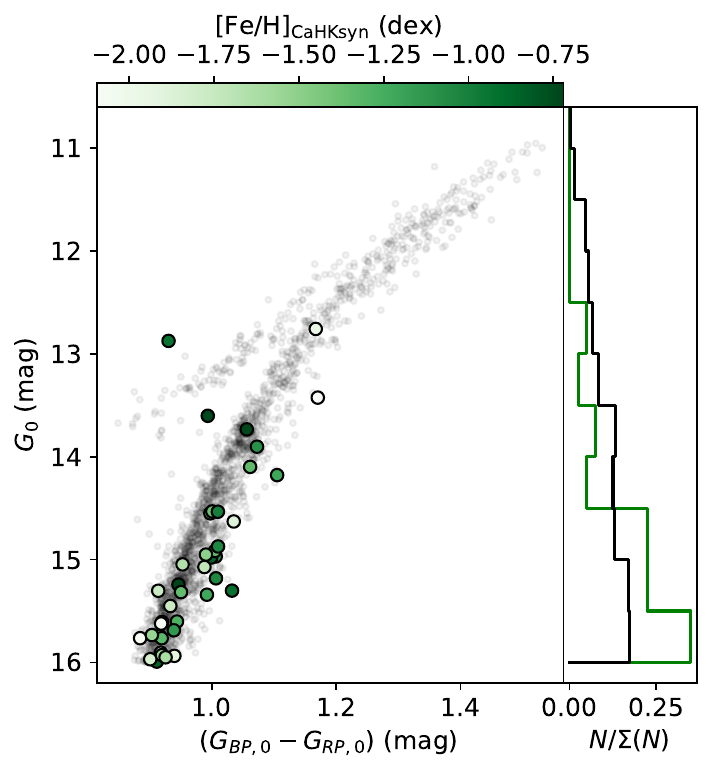}
    \caption{Left: CMD of our conservative \wcen$\,$ sample stars overlaid on the {\it cluster} sample. Each point is coloured by its \FeH$_\textrm{CaHKsyn}$. They all follow the broad width of the RGB of \wcen. Right: Histogram of the CMD representing a normalised Luminosity function with the conservative sample represented in green, and the {\it cluster} sample in black. The majority of stars are towards the fainter photometric limit in the conservative sample, forming an excess towards the faint end and a lack of stars towards the bright end of the RGB. This is consistent with the notion that the lower mass stars are preferentially ejected in GC disruption.}
    \label{fig:CMD_conservative}
\end{figure}

\begin{figure}
    \centering
    \includegraphics[width=1\columnwidth]{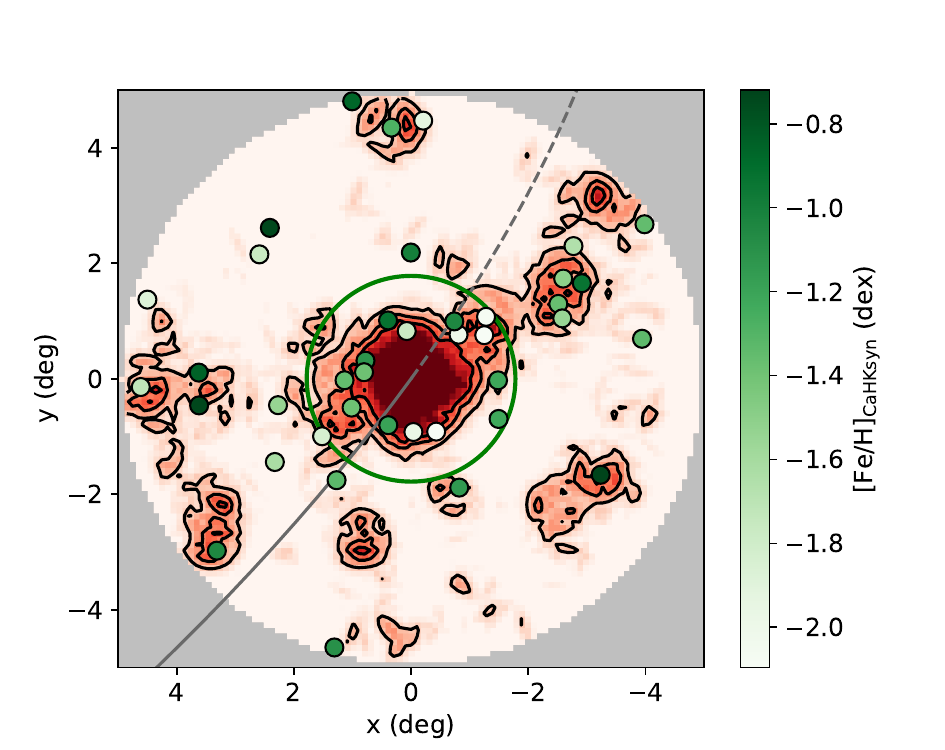}
    \caption{The same as Fig. \ref{fig:2d_prob}, with the addition of the conservative ($P_{\mathrm{mem}}>0.99$) sample of stars outside the tidal radius. Each point has been coloured by their metallicity, increasing in saturation. The stars are distributed along the orbit and in the direction of the stream presented in \citet[][]{2021MNRAS.507.1127K}. The stars are well mixed along the stream, evidence that the tails are created from tidally stripped stars from \wcen.}
    \label{fig:2d_conserve}
\end{figure}

\begin{figure*}
    \centering
    \includegraphics[width=0.9\textwidth]{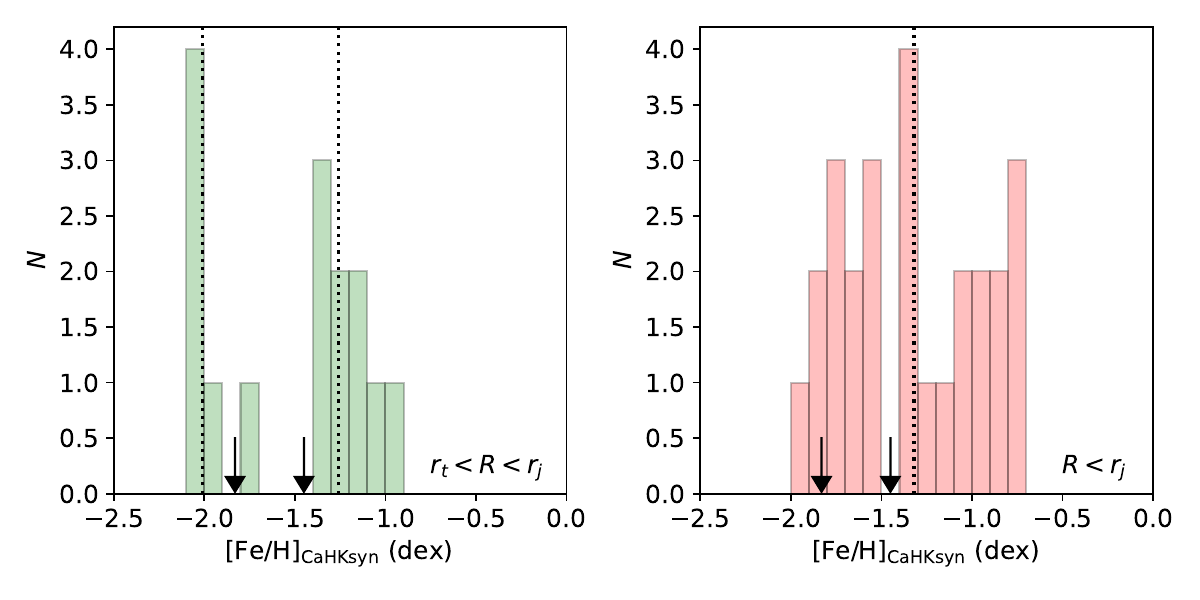}
    \caption{Histogram of our conservative sample between $r_{t}$ and $r_{j}$ (left) and beyond the Jacobi radius (right). The arrows indicate the measured peaks of the metallicity distributions between 15 arcmin and $r_t$ (right plot of Fig. \ref{fig:1d_hist_CL}), -1.83 and -1.45 dex. The vertical dotted demonstrates the mean \FeH$_\textrm{CaHKsyn}$: -2.01 and -1.23 dex on the right, -1.40 dex for the left. The two distinct populations are obvious in the regions between $r_t$ and $r_j$, while the broad population follows the same \FeH$_\textrm{CaHKsyn}$ range as the main body.}
    \label{fig:feh_hist}
\end{figure*}

The stellar populations with the tidal radius of \wcen$\,$have been studied quite extensively, but as we move to larger clustercentric radii, the picture is less well known. In Fig. \ref{fig:R_V_FeH}, we present the metallicity distribution as a function of radius for all stars in the sample. The field distribution is displayed in the lower plot of Fig. \ref{fig:R_V_FeH}, using stars with \Pmem $< 0.1$, and the top plot shows the \Pmem $> 0.1$ stars. The top row shows a large number of stars outside the tidal radius (denoted as the leftmost vertical dashed line in Fig. \ref{fig:R_V_FeH}) that have probabilities $P_{mem}>0.1$. However, the lower plot displaying the field shows that we may still have some level of contamination in stars with \Pmem $> 0.1$, particularly at for stars with \FeH $>-1.0$ dex. Therefore, to comment on the metallicities along the debris, we adopted a very conservative probability threshold to analyse, \Pmem $> 0.99$. These stars are demonstrated with large white circles in Fig. \ref{fig:R_V_FeH}, and in Figures \ref{fig:CMD_conservative} and \ref{fig:2d_conserve}, we show their distribution on the \wcen$\,${\it cluster} sample CMD and on the 2D surface density map presented in Fig. \ref{fig:2d_prob} respectively. We can see that there is still a decent number of stars that contain $\textrm{[Fe/H]}_\textrm{CaHKsyn}$ values that match the populations seen inside the main body and that they are distributed about the tidal tails in a manner that follows the tails and orbit. In Fig. \ref{fig:CMD_conservative}, we show that the conservative sample of stars follows the broad nature of the \wcen$\,$RGB. Additionally, the majority of the stars are towards the faint limit of our selection, as demonstrated by the normalised histogram representing the luminosity function of the conservative sample and the {\it cluster} sample in the right figure of \ref{fig:CMD_conservative}. This is to be expected, considering that the under theory of GC disruption, the lower-mass stars are preferentially lost \citep[e.g.,][]{2010MNRAS.401..105K}. If we consider the $\sim$55 arcmin region between the tidal radius and Jacobi radius, we see two distinct populations of stars in Fig. \ref{fig:R_V_FeH}. We display this separation in the histograms, created in the same manner as Fig. \ref{fig:1d_hist_CL}, in Fig. \ref{fig:feh_hist} but with a bin size of 0.1 dex. Between the tidal and Jacobi radius, we measure one metal-poor group with $\textrm{[Fe/H]}_\textrm{CaHKsyn} = -2.01 \pm 0.07$ dex and the more metal-rich group measured a $\textrm{[Fe/H]}_\textrm{CaHKsyn} = -1.26 \pm 0.06$ dex. These two groups are consistent with populations within the central regions of \wcen. Considering all stars within this range, we find a mean metallicity of $\textrm{[Fe/H]}_\textrm{CaHKsyn} = -1.50 \pm 0.12$, and a dispersion of $\sigma_{\textrm{[Fe/H]}_\textrm{CaHKsyn}}=0.32\pm0.01$. On average, these are very similar to the mean and spread of central regions. This work is the first detection of multiple stellar populations beyond the tidal radius of \wcen. Further, these stars are distributed about the cluster centre in a mixed manner: neither extension has an overabundance of stars from either group out to the Jacobi radius (the green ring in Fig. \ref{fig:2d_conserve}).

Increasing the clustercentric radius, Fig. \ref{fig:R_V_FeH} shows that the metallicities retain their broad distribution like what is seen within the tidal radius. Over 3 degrees in radius, we find the mean metallicity to be $-1.40 \pm 0.06$ dex (right histogram in Fig. \ref{fig:feh_hist}), which increases to $-1.32$ dex when incorporating stars between the tidal and Jacobi radius. This average value is entirely consistent with the metal-moderate populations in $\omega$ Cen and expresses a metallicity dispersion of $\sigma_{\textrm\FeH_{\textrm{CaHKsyn}}}=0.34\pm0.06$ - similar to the stars in the outer regions of \wcen. Further, beyond the Jacobi radius, the populations retain their well-mixed nature as Fig. \ref{fig:2d_conserve} demonstrates, much like the outer regions of the cluster. A few stars are found off-stream, but that does not discount them as potential members, as individual stars of \wcen$\,$have been identified throughout the MW halo \citep[][]{2020MNRAS.491.3374S,2023MNRAS.524.2630Y}. By considering all the factors discussed, the peripheral structure we are identifying here is highly likely to be constructed of tidally stripped stars from within the tidal radius of \wcen.

\section{Discussion}
\subsection{Consistency in the Inner Regions}\label{sec:Disc_1}
\begin{figure*}
    \centering
    \includegraphics[width=\textwidth]{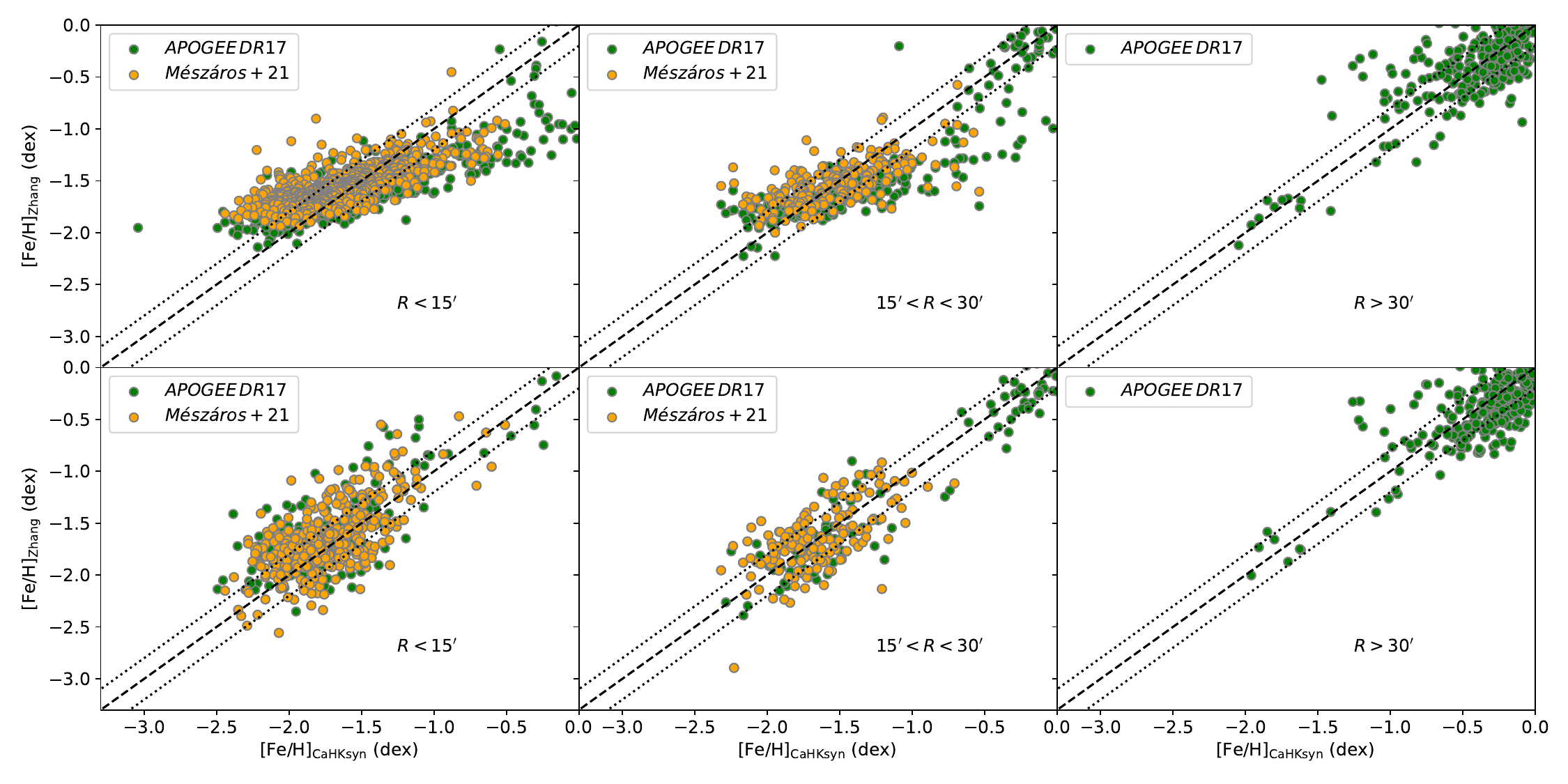}
    \caption{Demonstration of the metallicity relationship of different surveys and metallicities scales. Top: Relationship between Pristine-Gaia-synthetic catalogue, and both \citet[][]{2021MNRAS.505.1645M} and APOGEE DR17 \citep[][]{2020AJ....160..120J}. Bottom: Same stars on the top row, but now between Pristine-Gaia-synthetic and \citet[][]{2023MNRAS.524.1855Z}. Left: All stars within 15 arcmin. Middle: All stars between 15 and 30 arcmin. Right: All stars from 30 arcmins to 2 degrees. The dashed line indicates a one-to-one relationship, with the dotted lines showing a 0.2 dex offset. In the top row, the stars within 15 arcmin do not match one-to-one on the two ends of the GC metallicity distribution, and the effect is less pronounced in the middle figure. Outside 30 arcmin the relationship between APOGEE DR17 and the Pristine-Gaia-synthetic catalogue is one-to-one. The one-to-one relationship developing as a function of radius implies completeness is affecting the synthetic photometry. The bottom row shows a one-to-one correspondence between Pristine-Gaia-synthetic and \citet[][]{2023MNRAS.524.1855Z}, indicating a common origin of the \FeH$\,$relationship seen in the top row.}
    \label{fig:FeH_Mez}
\end{figure*}

The \wcen$\,$inner regions we explore overlap with spectroscopic surveys and works, presenting an opportunity to compare the metallicities. We cross-matched our data with the $\sim800$ red giants in-common from M21, which utilized the BACCHUS code on APOGEE DR16 \cite[][]{2020ApJS..249....3A,2020AJ....160..120J} spectra to measure abundances, and $\sim1500$ stars from APOGEE DR17 \citep[hereafter APOGEE][]{2022ApJS..259...35A}. Additionally, we cross-matched those two data sets with \citet[][hereafter Zhang]{2023MNRAS.524.1855Z}, whose [Fe/H] measurements are derived from data-driven modelling of Gaia XP spectra. We present a comparison of the M21 and APOGEE metallicities with this work and \citet[][]{2023MNRAS.524.1855Z} in Fig. \ref{fig:FeH_Mez}, sorting the stars into three radial bins, within 15 arcmin, between 15 and 30 arcmin, and outside 30 arcmin, which includes those in APOGEE to a radius of 5 degrees. While the extensive calibration of Pristine shows clear one-to-one relationships with other high-resolution spectroscopy (see Fig. 12 in \citealt[][]{2023arXiv230801344M}) including APOGEE, Fig. \ref{fig:FeH_Mez} does not show the same relationship for the stars in common between Pristine and M21 within 15 arcmins, and to a lesser extent between 15 and 30 arcmin. We find that the Pristine-Gaia-synthetic catalogue typically provides a higher metal content for stars with $\textrm{[Fe/H]}_\textrm{CaHKsyn} > -1.0$ dex, and a lower [Fe/H] value for $\textrm{[Fe/H]}_\textrm{CaHKsyn} < -2.0$ dex when compared to M21 and APOGEE. \citet[][]{2023arXiv230801344M} states that the Pristine methodology is primed to explore stars with $\textrm{[Fe/H]}_\textrm{CaHKsyn} < -1.0$ dex by design, and issues may arise when exploring the Pristine data set for stars with higher metallicities. This could be what we are seeing in Fig. \ref{fig:FeH_Mez} in the $\textrm{[Fe/H]}_\textrm{CaHKsyn} > -1.0$ regime. Interestingly, the relationship between the Zhang and Pristine-Gaia-synthetic catalogues does appear to be one-to-one across all three regions. By extension, Zhang would also show the same relationship if we compared their metallicities to M21 and APOGEE. This implies that there may be a Gaia-related issue affecting the metallicity measurements in the central region of \wcen, most likely crowding affecting the XP spectra. The fact that the relationship between the M21 and APOGEE datasets with Pristine-Gaia-synthetic catalogues becomes more one-to-one at increasing clustercentric radius adds evidence for crowding being the culprit. Therefore, while we present the full Gaia-Pristine-synthetic metallicity distribution of \wcen$\,$here, we do stress that the metal-poor populations at -2.0 dex and below may be more metal-rich that reported in the Pristine-Gaia-synthetic catalogue for the stars within 15 arcmin.

\subsection{From the Cluster to the Streams}
Establishing a connection based on metallicity between tidally stripped stars and their progenitors has always been difficult. The suggestion made in the previous section that the extended structure identified$\,$in the periphery of \wcen$\,$originates from the main body has potential implications for the halo substructures that are suggested to be related to \wcen$\,$in some way. While there have been many connections of substructure to \wcen$\,$in the past decade or so, there has been little chemical tagging. The detection of multiple stellar populations in the stars in the outer regions of \wcen, may give insight into the populations being excited into the outer regions and those being stripped. A study of MUSE spectroscopy by \citet[][]{2024ApJ...970..152N} has shown that central regions of \wcen$\,$ are well mixed (i.e., populations show no distinct radial or spatial distribution) out to the half-light radius. That finding is supported by the spectroscopic survey of \citet[][]{2020AJ....159..254J}, who show that the cluster is well mixed out to a clustercentric radius of approximately 20 arcmin by exploring the metallicity distribution as a function of radius, where the vast majority of their analysed stars lie. Though they did demonstrate that the most metal-poor populations of \wcen$\,$($-2.4 < $ \FeH$_\textrm{CaHKsyn}$ $ < -2.1$ dex) are centralised. One of the larger photometric studies of \wcen$\,$performed with DECam by \citet[][]{2017AJ....153..175C} showed that the more-metal rich populations are less centrally populated in \wcen, out to a distance of 30 arcmins. Beyond 30 arcmins, \citet[][]{2017AJ....153..175C} found that the cluster appears to be well mixed as well. Between $r_t$ and $r_j$, we observe that one of the two populations seen in Figures \ref{fig:R_V_FeH} is the metal-rich population. This is an interesting finding, especially when considering the previously reported observation that the more metal-rich population is more spatially extended. Additionally, we can see the presence of the metal-poor population in the same region, which is evident in the well-mixed central regions and continues into the periphery. Beyond the Jacobi radius, the tidal tails also appear to be well mixed. If the tidal tails consist of directly stripped stars from \wcen, we would expect them to share a similar stellar population. Interestingly, the most common populations of \wcen$\,$(between -1.8 to -1.4 dex) appear not to be present in the region between $r_t$ and $r_j$, but are present in the conservative sample in the tidal tails. On this matter, we mentioned in the previous section that Fig. \ref{fig:CMD_conservative} shows that most of the stars detected are towards the faint limit of Pristine. \wcen$\,$does have a significant population of stars at these magnitudes as indicated by many studies of Gaia, therefore deeper photometry/targeted spectroscopy should fill in that gap between the populations in the central regions and tidal regions. Ultimately, however, the combination of spectroscopy with deeper photometry would create a larger sample of \wcen$\,$ to see if there is a lack of stars with intermediate metallicity in this region.

%The {\it Fimbuthal} tidal stream, when discovered by \citet[][]{2019NatAs.tmp..258I}, was tentatively connected to \wcen$\,$through stars identified with metallicities ranging from $-1.80$ to $-1.36$ dex. Subsequent works such as \citet[][]{2020MNRAS.491.3374S} identified two stars with GALAH DR3 \citep[][]{2015MNRAS.449.2604D, 2022MNRAS.510.2407B}, again with consistent metallicities ([Fe/H] = -1.88 and -1.53 dex) but also with consistent s-process heavy element abundances for [Y/Fe] and [Ba/Fe]. The unsupervised learning technique of \citet[][]{2023MNRAS.524.2630Y}, which utilises GALAH with Gaia EDR3, uncovered 18 more potentially stripped tidal stars but all these lacked that crucial connection to the stripped stars in the immediate environment. Our findings of a range of metallicities between -2.4 to -1.0 dex of the stars from the tidal radius into the tidal tails present a consistency between this work and those previously discussed. The resulting implication is that these stars all linked with \wcen$\,$may very well be members - they have consistent metallicities with the tidally stripped stars. The connection to {\it Stream $\#$55} identified as part of the trailing arm of \wcen, uncovered in \citet[][]{2024ApJ...967...89I}, is based on kinematics/action energy space. \citet[][]{2024ApJ...967...89I} does not provide metallicities for {\it Stream $\#$55}, and therefore we can not comment on the chemical connection of {\it Stream $\#$55} to the stars in the tidal tails studying here.

Beyond just connecting \wcen$\,$to its tidally stripped stars, the curiosity of the cluster being the core of a now-defunct dwarf galaxy deserves some attention in the context of the stars in the tidal tails close to the cluster. Numerous works over the past few years have linked \wcen$\,$to different accretion events. Primarily, there are two events that \wcen$\,$is proposed to be a part of: the GSE dwarf galaxy, and the Sequoia dwarf galaxy. However, the metallicity distributions of GSE, Sequoia, and \wcen$\,$ do not demonstrate which dwarf galaxy \wcen$\,$belongs to. \citet[][]{2021MNRAS.508.1489F} and \citet[][]{2022ApJ...935..109L} presented two metallicities distributions of GSE and Sequoia stars and both dwarf galaxies have distributions that are overall more metal-rich than \wcen. with peaks of the distribution place Sequoia at $\sim -1.3$ dex and GSE are at a more metal-rich at $\sim-1.14$ dex, when compared to the major peaks of -1.83 and -1.50 dex of this work. There is a metal-poor tail in both GSE and Sequoias metallicity distribution towards \FeH$\,= -2.0$ dex, but there is not enough data to suggest whether \wcen$\,$belongs to GSE or Sequoia based on chemistry. Ultimately, chemistry and line-of-sight kinematics are required to comprehend the nature of extra-tidal structure fully. This will become possible with the next generation of telescopes and upcoming spectroscopy surveys. The multi-object capabilities of 4MOST \citep[][]{2019Msngr.175....3D,2023Msngr.190...13L} will be able to explore \wcen$\,$efficiently; to gather the line-of-sight kinematics required to fully explore any connection with dwarf galaxy accretion events to \wcen$\,$and its tidal tails, and other MW GCs. Additionally, other upcoming instruments and surveys such as WEAVE \citep[][]{2012SPIE.8446E..0PD} on the WHT, MOONS \citep[][]{2020Msngr.180...10C} on the VLT, and the Prime Field Spectrograph \citep[PFS;][]{2016SPIE.9908E..1MT} on the Subaru telescope will all revolutionise our view of GCs. We look forward to new findings of these new instruments that further explore the connections of GCs to accretion events or tidal streams, by looking at the poorly understood region in the immediate periphery of MW GCs.

\section{Conclusion}
The Pristine survey provides synthetic CaHK photometry across most of the night sky, allowing for an extra dimension in the analysis of GC peripheries. We have explored \wcen$\,$and its periphery with Pristine, identifying \wcen$\,$stars based on Gaia astrometry, Gaia CMD locale and colour-colour space defined by $\rm{(CaHK_{ind}, (G_{BP,0}-G_{RP,0})})$. Our probabilistic approach assigned each star in a 5-degree clustercentric radius a value between 0 and 1, representing the probability that a given star is a \wcen-like star. We recover the tidal tails of \wcen, and assess the metallicity distribution of \wcen$\,$as a function of clustercentric radius. Within 15 arcmin, we find peaks in the metallicity distribution at $-2.11 \pm 0.03$, $-1.83\pm 0.02$, $-1.50\pm 0.02$ and $-1.22 \pm 0.06$ dex, and between 15 arcmin and the tidal radius, we find two main populations of $-1.82 \pm 0.03$ and $-1.45 \pm 0.04$ dex, consistent with the most dominate populations present within 15 arcmin of the cluster center. These populations are in firm agreement with prior works. However, we warn that the populations that are more metal-poor than -2.0 dex within the cluster may not be as metal-poor as reported, likely due to the effects of crowding. Between the tidal radius and Jacobi radius, we uncover the first detection of multiple stellar populations in \wcen$\,$ outer regions, detecting mean metallicities at $-2.01$ and $-1.26$ dex which are comparative to a couple of the peaks in the \wcen$\,$ metallicity distribution. In the periphery/extended structure, we find a broad metallicity distribution that closely resembles populations within the cluster, the first evidence of multiple stellar populations in tidal tails from a presently disrupting GC. We note that the stars of broad metallicities are found populated throughout the tail, implying that the tails are mixed, much like the cluster. This result reinforces the tails are indeed the result of tidal disruption of \wcen.

Further into the Galactic halo, we show that stars stripped can be traced back to \wcen$\,$and its periphery. The tidal streams {\it Fimbuthal} and {\it Stream $\#$55} connection to \wcen$\,$ can be tested through the chemistry of the stars that are being lost, and indeed the metallicities of the small handful of stars identified in the {\it Fimbuthal} stream in \citet[][]{2019NatAs.tmp..258I} are found to be consistent with this new selection of extra-tidal stars. The stars in \citet[][]{2012MNRAS.427.1153S} and \citet[][]{2023MNRAS.524.2630Y} are comparable as well. This result agrees with the conclusions of those works that the origin of all these substructures and individually tagged stars are tidally stripped from \wcen. 

This work demonstrates the need for chemistry in extra tidal regions, or even the outermost regions, of GCs to explore the role GCs play in the build-up of the MW. The metallicities of Pristine are a crucial addition to identifying extra tidal stars. For the first time, we can make general statements about the metallicity distributions at large clustercentric radii. The lack of line-of-sight velocities does hinder our ability to create the cleanest of \wcen$\,$stars in the outer regions and beyond, but the next generation of multi-object spectrographs will be able to provide that crucial data to deeper photometric depths where the majority of tidally stripped stars will be. We envisage that more extra tidal stars will be identified, and any host GCs connection to any one of the plethora of tidal streams and galactic accretion events will be solidified. In turn, we can paint the clearest picture of the Milky Way substructure and its build-up.

\section*{Acknowledgements}
This work makes use of the following software packages: \textsc{astropy} \citep{2013A&A...558A..33A,2018AJ....156..123A}, \textsc{astroquery} \citep{2019AJ....157...98G}, \textsc{Gala} \citep{gala,adrian_price_whelan_2020_4159870}, \textsc{matplotlib} \citep{Hunter:2007}, \textsc{numpy} \citep{2011CSE....13b..22V}, \textsc{scipy} \citep{2020SciPy-NMeth}, \textsc{sklearn} \citep[][]{scikit-learn}, \textsc{Ultranest} \citep{2014A&A...564A.125B,2019PASP..131j8005B,2021JOSS....6.3001B}.
We thank the anonymous referee for their comments which improved the quality of this paper. The authors thank Annette Ferguson for useful discussions that improved the quality of this paper. PBK acknowledges support from the Japan Society for the Promotion of Science under the programme Postdoctoral Fellowships for Research in Japan (Standard). This work was supported by JSPS KAKENHI Grant Numbers JP23KF0290,  JP21H04499, JP20H05855. 

This work has made use of data from the European Space Agency (ESA) mission
{\it Gaia} (\url{https://www.cosmos.esa.int/gaia}), processed by the {\it Gaia}
Data Processing and Analysis Consortium (DPAC,
\url{https://www.cosmos.esa.int/web/gaia/dpac/consortium}). Funding for the DPAC
has been provided by national institutions, in particular the institutions
participating in the {\it Gaia} Multilateral Agreement.

%%%%%%%%%%%%%%%%%%%%%%%%%%%%%%%%%%%%%%%%%%%%%%%%%%
\section*{Data Availability}
The data sets that are unpinning this paper are available at the Pristine official website. Additionally, the membership probabilities of all the stars in this paper are available online at DOI: 10.5281/zenodo.14791430.



\bibliographystyle{mnras}
\bibliography{Library_bibtex.bib}

% Don't change these lines
\bsp	% typesetting comment
\label{lastpage}
\end{document}